\documentclass[preprint2]{aastex}

\def \etal{\mbox{et al.}}

\slugcomment{Millenium Essay}

\shorttitle{GALAXY EVOLUTION}
\shortauthors{ELLIS}

\begin{document}


\title{Cross-Roads in Studies of Galaxy Evolution}


\author{RICHARD S. ELLIS\altaffilmark{1}}
\affil{Astronomy Department, MS 105-24, Caltech, Pasadena CA 91125; rse@astro.caltech.edu}

\altaffiltext{1}{This Essay is one of a series of invited
contributions which will appear in the {\em PASP} throughout
2000-2001 to mark the upcoming millenium (Eds.)}

\begin{abstract}
Studies of galaxy evolution at optical and near-infrared
wavelengths have reached an interesting point in their historical
development and the arrival of a new millenium provides an
appropriate occasion to review the overall direction in which this
subject is moving. The aim of this short essay is to emphasize the
qualitatively different datasets now arriving and the more
detailed and challenging scientific analyses needed for future
progress.

\vspace*{40pt}
\end{abstract}


\section{Introduction}

Progress made in studying faint galaxies is no longer measured
solely in terms of the depth at which samples are located but also
in terms of the ever-increasing detail provided for objects at
fixed redshifts and flux limits. Consider a typical 23rd magnitude
galaxy at a redshift $z\simeq$0.7. In the 1970's our knowledge
about such an object would be very limited; it would have a
photographic apparent magnitude and possibly a single broad-band
color. By the mid-1980's, it might have some CCD photometry and a
low dispersion spectrum from a 4-meter telescope yielding a
redshift, k-corrected luminosity and some emission or absorption
line characteristics. By the mid-1990's a WFPC-2 image from Hubble
Space Telescope (HST) would provide morphological and structural
information, and ground-based photometry over a range of
wavelengths from the UV to the IR might characterise its spectral
energy distribution (SED) fairly well. Late in the 1990's,
intermediate dispersion spectroscopy with the new generation of
8-10 meter telescopes has provided some dynamical data, in terms
of line widths or, exceptionally, rotation curves. This trend of
increasing detail on high redshift galaxies is sure to continue as
we anticipate obtaining {\em resolved spectra} from integral field
unit spectrographs on ground-based telescopes with adaptive optic
feeds, or using the Next Generation Space Telescope.

How is the community responding to the challenge of interpreting
these increasingly-detailed datasets and to what extent does the
emerging picture of how galaxies evolve indicate new directions
which define future instrumental requirements? The question is not
merely of sociological interest. Strategic changes in direction,
when necessary, represent a natural outcome of the scientific
process and thus it is important to try and anticipate them. This
is particularly so for faint galaxy studies which continue to
motivate the construction of future large telescopes.

Until recently, observers of faint galaxies were largely
preoccupied with conducting `census surveys', whose aim is to
define the epochs within which galaxies of various photometric
selections lie, as a precursor to evolutionary tests. We may be
approaching the time when such studies will be complete and thus
face something of a ``cross-roads" in the subject. In this essay I
discuss whether both observers and theorists are ready to
interpret the much more detailed datasets that we can expect
fairly soon.

\section{Earlier Cross-Roads}

We have encountered cross-roads before in our studies of galaxy
evolution. A major one occurred when it was realised that main
sequence brightening arising from stellar evolution in
first-ranked cluster galaxies could significantly complicate their
use in tracing the history of the cosmic expansion. That distant
galaxies might be useful as probes of galactic history did not, of
course, escape the attention of early pioneers \citep{sandage61}.
However, the (still-familiar) eagerness to constrain the
cosmological parameters clearly took priority. Tinsley was perhaps
the most ardent advocate in recognizing this cross-roads
\citep{tinsley76} convincingly inspiring observers to return to
studies of the counts and colors of faint field galaxies with the
new generation of wide-field 4-meter telescopes
\citep{bap79,tyson79,kron80,koo85}. The emerging data were
compared with her evolutionary synthesis models
\citep{tinsley77,tinsley80} and laid the foundation for an entire
industry of surveying the faint skies which has grown apace since
1980.

In present-day parlance, Tinsley assumed galaxies underwent ``pure
luminosity evolution''. They evolved in isolation and thus were
conserved in number. Different star formation histories were
chosen to reproduce the colors of the present-day Hubble sequence
in an admirably simple scheme \citep{struckmarcell78}. As noted by
earlier workers, the present colors of spheroids can be fit with
model galaxies with a high formation redshift and a rapidly
declining star formation rate; such objects would be spectacularly
luminous and blue close to their formation epoch. The faint blue
excess seen in the counts \citep{rse97} could originate from such
a primordial population. Finding primeval galaxies and discovering
the {\em epoch of galaxy formation} spectroscopically thus became
a new `Holy Grail' for observers. For many reasons, these searches
were largely inconclusive \citep{pritchet94} but they did serve to
emphasize the importance of faint object spectroscopy.

Efficient spectrographs equipped with multi-object feeds enabled
systematic surveys of magnitude-limited samples heralding
importance advances. Evolution of galaxy populations could be
addressed by comparing well-defined samples within different
redshift intervals \citep{lilly95,rse96,cowie96,cohen00}. These
surveys have revealed a complex picture of luminosity-dependent
evolutionary trends with some evidence for an increase in number
density with look-back time.

A less obvious cross-roads appeared after the completion of the
10-meter Keck telescopes. It might seem obvious that more powerful
telescopes should be used to extend the earlier magnitude-limited
redshift surveys faintward. However, the mean redshift depth is
only marginally increased with a larger telescope in this way.
Most fainter galaxies are simply less luminous sources occupying
the same redshift space as those which dominate the 4-meter
surveys. Photometric selection techniques, the most successful of
which are based on Lyman and Balmer discontinuities, more
efficiently yield high redshift samples with surface densities
well-matched to Keck's LRIS \citep{steidel96,steidel00}. Such
techniques provided the first glimpse at the abundance of early
galaxies as well as their likely contribution to the star
formation and metal production history of the Universe
\citep{steidel99,madau99}.

\section{New Directions for High Redshift Studies}

So what can we learn from this brief history in terms of how
observational strategies might change in the future? Whilst
determining the redshift distribution of flux-limited samples,
$N(m,color,z)$, with survey spectrographs was successful in the
past, it is less likely to be a productive technique in the
future. Just as microwave background enthusiasts argue that
``soon" we will know the gamut of cosmological parameters (and so
can concentrate on understanding structure formation and the
nature of $\Lambda$ and dark matter) so, many argue, we may soon
have the statistical redshift distribution $N(z)$ for any
population we desire from {\em photometric redshift techniques}
\citep{connolly96} rendering traditional redshift surveys
unnecessary, at least for the purpose discussed above.

This is not to underestimate the challenges that remain in
convincing skeptics of this optimistic viewpoint. Photometric
redshift techniques suffer from a number of potentially dangerous
pitfalls not least the reliance on template SEDs based either on
(i) local galaxy data (which surely cannot be representative of
the bewildering variety of objects we see at all epochs) or (ii)
evolutionary synthesis models which are largely untested except in
the bright regime where normal Hubble types still dominate. One
must also remember that the photometric precision appropriate in
the oft-quoted Hubble Deep Field comparisons \citep{hogg98} is
considerably better than that available in typical ground-based
imaging surveys.

Two important new scientific directions have emerged in the past
few years. First, photometric redshift selection can already
isolate samples sufficient to permit their angular clustering to
indicate the spatial correlation function at early times
\citep{adelberger98,daddi00,brunner00,mccarthy00}. This provides a
valuable measure of large scale structure intermediate in epoch
between the microwave background and the present, crucial for a
direct test of gravitational instability. Spectroscopic data is
still highly desirable in these imaging surveys because if,
ultimately, galaxy clustering can be characterised additionally by
dynamical or stellar mass, rather than via mere population colors
or star formation rates, the connection with contemporary model
predictions will be much sounder. These deep surveys differ
significantly from their local magnitude-limited equivalents (2dF,
SDSS) in that redshift-dependent selection effects complicate the
analysis necessitating forward modelling of the various biases
\citep{kauffmann99}.

The other direction of importance is one that was established via
the arrival of WFPC-2. The first resolved images of distant
galaxies caused much optimism in the subject because of the
potential for examining evolution of galaxy morphologies
\citep{griffiths94}. Most would agree that the recognition of
familiar-looking spheroidal and spiral galaxies at redshifts
$z\simeq$1 was an important development. However, as we anticipate
building ground-based instruments that will deliver resolved {\em
spectroscopy} of faint galaxies, it is relevant to ask what, in
terms of quantitative astrophysics, we can learn about galaxy
evolution from resolved data.

\section{The Challenges of Interpreting Resolved Galaxy Data}

I believe we face another significant cross-roads in the subject
of galaxy evolution. Resolved data on distant galaxies may soon be
arriving from a variety of facilities, ground-based and
space-based, but are we ready to deal with it?

We might begin addressing this problem by considering what has
been learnt already from the community's experience so far with
HST images of distant galaxies. The most significant conclusions
here have come from the very limited subset of HST imaging for
which ground-based redshift data is available. The mismatch in the
field of view between ground-based survey spectrographs and WFPC-2
has seriously limited progress in this regard and the arrival of
the Advanced Camera for Surveys will not make a big impact unless
there is a concerted campaign for panoramic imaging with HST.
Disregarding rich clusters, only the Groth strip, the
CFRS/LDSS/Hawaii survey and HDF fields have been mosaiced with HST
to match ground-based data. Collectively these represent a
pitifully small coverage considering the potential impact of HST
imaging in this subject area.

The primary scientific results on galaxy evolution from the
limited HST datasets include claims for (i) physically smaller
sources at high redshift \citep{pascarelle95, bouwens98}, (ii) an
increasing abundance of peculiar, possibly merging, blue
star-forming systems by z$\simeq$1
\citep{glazebrook95,driver95,rse00}, and (iii) a slow growth in
spiral disks \citep{lilly98} and field ellipticals
\citep{menanteau00,im00} over 0$<z<$1. Each is a purely empirical
result whose place in a quantative understanding of galaxy
evolution is hindered by, at best, a loose interface with
contemporary theories of structure formation. Whereas uncertain
assumptions concerning gas cooling, feedback and related star
formation physics are necessary to use structure formation models
to predict cosmic star formation histories, luminosities and color
distributions \cite{baugh98,cole00}, a more profound challenge is
now faced in extending theoretical models to predict detailed
morphological and structural information. In this respect, relying
solely on a foundation of gravitational physics may be quite
inadequate. One suspects the subject would benefit considerably
from the experience of those studying the detailed dynamics and
interstellar media of nearby systems. Some progress is possible
with dynamical data such as rotation curves and internal velocity
dispersions \citep{vogt97,mo98,treu00} but only a small subset of
the faint galaxy population can currently be studied in this way.
Indeed, where attempts have been made to mimic the hydrodynamical
evolution of galaxies, serious discrepancies arise, for example in
understanding how stellar angular momentum is sustained in disk
galaxies \cite{weil98}. Clearly, there is a lot of work for
theorists to do!

Moreover, the interpretational deficiency does not lie entirely at
the hands of the theorists. We can ask if observers have thought
hard enough how to optimally condense their resolved data into
useful astrophysical information. Ingenious ``objective" criteria
have been introduced to classify distant systems taking into
account instrumental, surface brightness and other,
redshift-dependent, biases \citep{abraham94,odewahn95,
abraham96,conselice00}. HST galaxy images can also be `decomposed'
structurally, yielding parameters such as disk/bulge ratios and
disk scale-lengths \citep{marleau98,ratnatunga99}, although the
practicality of doing this for anything other than familiar
regular systems remains unclear. Nonetheless, a gap remains
between any objective taxonomic classification and astrophysical
conclusions if, as seems likely, galaxies transform from one class
to another as a result of environmental or merging processes
\cite{dressler97}. Perhaps tools applicable to the entire
population, equivalent to the stellar Hertzsprung-Russell diagram,
are needed to make progress in terms of a bigger picture
\cite{rse00}.

The observational momentum towards the provision of detailed
information on high redshift galaxies is increasing, which serves
only to emphasize the need to grasp the challenge of its
interpretation. On the future instrumentation web sites for the
current suite of 8-10 meter telescopes, integral field unit
spectrographs represent a dominant feature\footnote{An integral
field unit remaps the 2-D image of a resolved object so that each
pixel yields an independent spectrum \cite{bacon00}}. Providing
resolved galaxy data over $2<z<5$ likewise forms a major driver
for multi-conjugate adaptive optics on current and future large
telescopes as well as for NGST instruments.

Resolved data for samples of high redshift galaxies is not only an
inevitable next step in the development of astronomical
instrumentation, but also a logical scientific direction if we
wish to truly understand the variety of galaxy types we see today.
However, observers and theorists need to invest far more
intellectual effort to reap the benefits of the data arriving
soon.




\end{document}